\documentclass[article]{revtex4}
\usepackage{graphicx}
\usepackage{amssymb}
\usepackage{bm}
\usepackage{hyperref}
\usepackage{epstopdf}
\usepackage{subfigure}

\begin{document}

\title{Phantom dark energy as an effect of bulk viscosity}

\author{Hermano Velten$^{1}$\email{velten@physik.uni-bielefeld.de}, Jiaxin Wang$^{2}$\email{jxw@mail.nankai.edu.cn}, and Xinhe Meng$^{2,3}$\email{xhm@nankai.edu.cn}}
\affiliation{$^1$Departamento de F\'sica, Universidade Federal do Esp\'irito Santo, Vit\'oria, 29060-900, Esp\'irito Santo, Brazil\\
$^2$Department of Physics, Nankai University, Tianjin 300071, China\\
$^3$Kavli Institute of Theoretical Physics China,CAS, Beijing 100190, China\\}

\date{\today}

\begin{abstract}
In a homogeneous and isotropic universe bulk viscosity is the unique viscous effect capable to modify the background dynamics. Effects like shear viscosity or heat conduction can only change the evolution of the perturbations. The existence of a bulk viscous pressure in a fluid, which in order to obey to the second law of thermodynamics is negative, reduces its effective pressure. We discuss in this study the degeneracy in bulk viscous cosmologies and address the possibility that phantom dark energy cosmology could be caused by the existence of non-equilibrium pressure in any one of the cosmic components. We establish the conditions under which either matter or radiation viscous cosmologies can be mapped into the phantom dark energy scenario with constraints from multiple observational data-sets.
\end{abstract}

\maketitle

\section{Introduction}

Viscous cosmological scenarios have been widely studied since the development of relativistic thermodynamics. In fact, in a real universe it is very unlikely that dissipative processes do not take place. Since the standard description of the large scale cosmic dynamics assumes the cosmological principle, directional phenomena like shear and heat conduction are excluded and therefore, in a homogeneous and isotropic background, only bulk viscosity is allowed for cosmic fluids.

In the Eckart frame \cite{eckart}, first order deviations from equilibrium are expressed in the energy-momentum tensor as an additional non-adiabatic contribution $\Delta  T^{\mu \nu}$ as
\begin{equation}
T^{\mu \nu}= \rho u^{\mu}u^{\nu}+p_{\rm k} \, h^{\mu\nu} +\Delta T^{\mu \nu}, \hspace{0.5cm} {\rm with} \hspace{0.5cm} \Delta T^{\mu \nu}=-\xi u^{\gamma}_{; \gamma} h^{\mu\nu},
\end{equation}
where $ h^{\mu\nu}=u^{\mu}u^{\nu}+g^{\mu\nu}$ and $\xi$ is the positive (due to the second thermodynamics law \cite{weinberg1}) coefficient of bulk viscosity. For the background, $u^{\gamma}_{; \gamma}=3\text{H}$ which means that in the presence of bulk viscosity the effective pressure of the cosmic fluid is composed by the kinetic pressure $p_{\rm k}$ and the bulk viscous one as
\begin{equation}
P_{\rm eff}=p_{\rm k}-3\text{H}\xi.
\end{equation}
Bulk viscosity is the unique effect in nature capable to reduce the kinetic pressure of a fluid.

The quantity $\Delta T^{\mu \nu}$ is constructed in such way that the conservation ($T^{\mu}_{\nu \, ; \; \mu}=0$) still holds in the presence of dissipative contributions. Moreover, standard cosmology assumes that the conservation of $T^{\mu \nu}$ holds separately for each cosmic component. Assuming a typical equation of state $P_{\rm eff}=w\rho$ one finds
\begin{equation}\label{conserv}
\dot{\rho}+3\text{H}\rho(1+w)=0.
\end{equation}
For standard adiabatic ($\xi=0$) cosmic fluids the equation of state parameter assumes the values $w_{\rm m}=0$ for matter (dark matter and baryons) and $w_{\rm r}=1/3$ for relativistic (neutrinos and photons). Solving (\ref{conserv}) for these fluids we find $\rho_{\rm m}\propto (1+z)^3$ and $\rho_{\rm r}\propto (1+z)^4$, respectively. Dark energy can also be described by $P_{\rm de}=w_{\rm de}\rho_{\rm de}$ where the cosmological constant $\Lambda$ is recovered if $w_{\rm de}=-1$. Thus, if a cosmic fluid has bulk viscous pressure the background dynamics is modified because the density evolution of such viscous fluid will be different.

There were remarkable cosmological applications of viscous imperfect fluids already in the 1970s \cite{IsraelV, Klimek, Murphy, Belinskii, weinberg}. Due to the fact that bulk viscous pressure is negative, an inflationary epoch driven by bulk viscous pressure has also been proposed in the 1980s \cite{Diosi, waga, barrow,gron}. All these works have analysed the role played by bulk viscosity in the early universe. However, much before the discovery of the accelerated expansion of the universe (the dark energy phenomena) in 1998, one can find some mentions for a late time viscous universe \cite{potuba, padmanabhan}. The dark energy phenomena as an effect of the bulk viscosity in the cosmic media has been first investigated in refs. \cite{zimdahl, balakin}. These works are pioneering papers on cosmological bulk viscosity, but it is also worth noting some recent applications of viscous fluids as candidates for dark matter \cite{viscousDM}, dark energy \cite{viscousDE} or unified models \cite{unified}, i.e., when a single substance plays the role of both DM and DE simultaneously. In general, these applications rely on the phenomenological ground only, but there are some attempts in the literature to justify it \cite{winfried1}. However, just to cite an example concerning the existence of bulk viscosity in the universe, it has been demonstrated a long time ago that a gas of neutrinos do have bulk viscous properties \cite{degroot}. Even so, it is quite surprising that analysis in the field of neutrino cosmology do not take bulk viscosity into account.

Now, let us introduce a second topic of interest in this paper which is phantom dark energy fluid. This is a exotic type of fluid whose equation of state parameter violates the null energy condition, i.e. $w_{\rm de}<-1$. But first, remember that the cosmological constant ($\Lambda$) is the simplest and the most efficient way to generate the observed accelerated background expansion within general relativity (GR). It is the main component of the $\Lambda$CDM model which, after the crossing of many different observational data, has emerged as the standard description for the cosmic evolution \cite{WMAP9CP,SloanCP}.

The effective $\Lambda$ (the quantity that appears in the Friedmann equations) is inferred from the observations while its theoretical value can be calculated from quantum field theory. Comparing these quantities we find $\Lambda^{th} / \Lambda^{obs}\sim 10^{72} {\rm GeV}^4 / 10^{-47} {\rm GeV}^4=10^{119}$ and it is easy to realize the so called {\it cosmological constant problem} \cite{LambdaProb}. Thus, the actual nature of the DE can not be $\Lambda$ (interpreting it as the quantum vacuum) though the observations say the contrary (see e.g., \cite{jerome} for a recent discussion). This motivates investigations for alternative dark energy models.

Assuming that standard fluids does not violate the null energy condition, equation (\ref{conserv}) can be used to set the constraint $w_{\rm de}\geq-1$. However, relaxing this condition, $w_{\rm de}$ is free to vary to values $w_{\rm de}<-1$ originating the idea of phantom fluid, i.e., an exotic component whose density increases as the Universe expands. This idea was pointed out for the first time in reference \cite{cadwell} and has the Big Rip scenario as a natural consequence \cite{BigRip}.

The phantom behaviour can also be constructed in terms of a scalar field with negative kinetic term 
\begin{equation}
\rho_{\rm de}=-\frac{1}{2}\dot{\phi}^2+V(\phi),\hspace{1cm} p_{\rm de}=-\frac{1}{2}\dot{\phi}^2-V(\phi) \hspace{1cm} \rightarrow \hspace{1cm} w_{\rm de}=\frac{-\frac{1}{2}\dot{\phi}^2-V(\phi)}{-\frac{1}{2}\dot{\phi}^2+V(\phi)},
\end{equation}
leading to the general constraint $w_{\rm de}<-1$ for $\dot{\phi}^2>0$.

Phantom dark energy models have been preferred by most of the observations \cite{Phantomdata}. Current notable results are $w_{\rm de}=-1.04^{+0.09}_{-0.10}$ \cite{nakamura}, $w_{\rm de}=-1.10^{+0.14}_{-0.14}$ \cite{komatsu} and the recent results from Planck satellite $w_{\rm de}=-1.49^{+0.65}_{-0.57}$ \cite{Planck2013CP}. Although the error bars cover the case $w_{\rm de}=-1$, it is worth noting a tendency for phantom values.

Our goal in this work is to bring together cosmological bulk viscous effects and phantom dark energy behaviour into the same discussion. We argue that the existence of a negative bulk viscous pressure in the description of ordinary cosmic fluids, e.g. radiation or matter, can lead to the phantom dark energy interpretation. The source of this issue relies on the assumption that different cosmic fluids obey separately the energy-momentum conservation whilst the full background dynamics is sensitive to the total pressure of the system rather than the individual pressures. Moreover, viscosity is very likely to occur in nature for the case of multi-fluid flows as seems to be the description of the Hubble expansion.

In the next section we describe particular viscous models which can be seen as phantom ones. In section 3 we use observational data to quantify this degeneracy. For example, allowing bulk viscous pressure in the radiation fluid we show how constraints on the radiation viscosity can be translated to constraints on the phatom dark energy equation of state. We also consider the case where matter has viscosity and the hypothetical situation where viscosity can be assigned to the vacuum energy. We conclude in section 4.

\section{Degeneracy in bulk viscous cosmologies}

The purpose of this section is to show that assigning bulk viscosity to either one specific cosmic component or to the total dynamics can lead to degenerated cosmological scenarios. This avoids the identification of the actual viscous fluid and possible leads to the phantom interpretation.

Our approach assumes general relativity 
\begin{equation}
R^{\mu \nu}-\frac{1}{2}g^{\mu \nu}R=8\pi G ~T_{\rm eff}^{\mu \nu},
\end{equation}
and a flat, homogeneous and isotropic expanding background. 

As a consequence of the Bianchi identities the total energy-momentum tensor is conserved $\partial_{\mu}T^{\mu\nu}=0$, then
\begin{equation}\label{Cont0}
\dot{\rho}_{\rm eff}+3\text{H}\left(\rho_{\rm eff}+P_{\rm eff}\right)=0 .
\end{equation}

Let us assume that bulk viscosity in the Universe is a collective phenomena acting on the global dynamics such that we rewrite (\ref{Cont0}) as

\begin{equation}\label{Cont}
\dot{\rho}_{\rm eff}+3\text{H}\left(\rho_{\rm eff}+P_{\rm per}+\Pi\right)=0,
\end{equation}
where the effective pressure is now decomposed into perfect-fluid pressure $p_{\rm per}$ and the viscous pressure, which in the Eckart formalism reads $\Pi=-3\text{H}\xi$. In order to solve for the dynamics of this system the next step is to set the bulk viscous coefficient $\xi$ which is a crucial aspect of any viscous fluid.

From kinetic theory the transport coefficients are calculated as a function of powers of the temperature of the fluid $\xi\equiv\xi(T)$ \cite{xiT}. Consequently, thermodynamics allows us to replace the temperature $T$ by the density which in our case is $\rho_{\rm eff}$. Then, following many works in the literature, we can write, via the Friedmann's equations, the coefficient of bulk viscosity as a function of the background expansion $\xi\equiv\xi(\text{H})$. 

Now, we turn our attention to the fact that the real Universe is made of different $i$ components. Assuming that $\rho_{\rm eff}=\rho_{\rm r}+\rho_{\rm m}+\rho_{\rm de}$ equation (\ref{Cont}) is then written as
\begin{equation}\label{ContT}
\dot{\rho}_{\rm m}+\dot{\rho}_{\rm r}+\dot{\rho}_{\rm de}+3\text{H}\left(\rho_{\rm m}+\frac{4}{3}\rho_{\rm r}+\rho_{\rm de}+p_{\rm de}-3\text{H}\xi\right)=0.
\end{equation}

The degeneracy appears in bulk viscous models with $\xi \equiv\xi(\text{H})$ because we have the freedom to further decompose the above equation mathematically either into the system
\begin{equation} \label{viscsm}
\dot{\rho}_{\rm m}+3\text{H}\left(\rho_{\rm m}-3\text{H}\xi\right)=0, \hspace{1cm} \dot{\rho}_{\rm r}+4\text{H}\rho_{\rm r}=0,\hspace{1cm} \dot{\rho}_{\rm de}+3\text{H}\left(\rho_{\rm de}+p_{\rm de}\right)=0;
\end{equation}
or into the system
\begin{equation}\label{viscsr}
\dot{\rho}_{\rm m}+3\text{H}\rho_{\rm m}=0, \hspace{1cm} \dot{\rho}_{\rm r}+3\text{H}\left(\frac{4}{3}\rho_{\rm r}-3\text{H}\xi\right)=0,\hspace{1cm} \dot{\rho}_{\rm de}+3\text{H}\left(\rho_{\rm de}+p_{\rm de}\right)=0;
\end{equation}
or even into
\begin{equation} \label{viscsl}
\dot{\rho}_{\rm m}+3\text{H}\rho_{\rm m}=0, \hspace{1cm} \dot{\rho}_{\rm r}+4\text{H}\rho_{\rm r}=0,\hspace{1cm} \dot{\rho}_{\rm de}+3\text{H}\left(\rho_{\rm de}+p_{\rm de}-3\text{H}\xi\right)=0.
\end{equation}

The choice for one of the above decompositions, either (\ref{viscsm}), or (\ref{viscsr}), or (\ref{viscsl}) corresponds to the cases where we assign viscosity to matter ($\xi_{\rm m}\equiv \xi_{\rm m}(\text{H})$), radiation ($\xi_{\rm r}\equiv \xi_{\rm r}(\text{H})$) or dark energy ($\xi_{\rm de}\equiv \xi_{\rm de}(\text{H})$), respectively. Now, note that we could have assumed the opposite starting point where only one cosmic component behaves as a viscous fluid with $\xi\equiv\xi(\text{H})$.

The above discussion is quite general and the decomposition of (\ref{ContT}) into one of the systems (\ref{viscsm}-\ref{viscsl}), or vice versa, is valid under the assumption $\xi\equiv\xi(\text{H})$. A different scenario appears if we restrict $\xi_i$ of the fluid (i) to be dependent on its own density, e.g. $\xi_{\rm m}\equiv \xi_{\rm m}(\rho_{\rm m})$, $\xi_{\rm r}\equiv \xi_{\rm r}(\rho_{\rm r})$ or $\xi_{\rm de}\equiv \xi_{\rm de}(\rho_{\rm de})$. In the next section wWe address the cases $\xi_{\rm i}\equiv \xi_{\rm i}(\rho_{\rm i})$, with ${{\rm i}={\rm m},{\rm r}}$.

\section{Phantom dark Energy via Bulk Viscosity}\label{sectwo}

\subsection{$w$CDM cosmology}
Let us first establish the standard $w$CDM dynamics in which the phantom dark energy appears. The total energy-momentum tensor is given by
\begin{equation}
T_{\left({\rm eff}\right)}^{\mu \nu}=T_{\left({\rm r}\right)}^{\mu \nu}+T_{\left({\rm m}\right)}^{\mu \nu}+T_{\left({\rm de}\right)}^{\mu \nu},
\end{equation}
where for the pressure of the fluids we use
\begin{equation}
P_{\rm eff}=p_{\rm r}+p_{\rm m}+p_{\rm de}=\frac{\rho_{\rm r}}{3}+w_{\rm de}\rho_{\rm de},
\end{equation}
where $w_{\rm de}$ is a constant parameter.
The background expansion for the $w$CDM model becomes
\begin{equation}\label{Ew}
E^{\rm w}(z)=\left[\Omega^{\rm w}_{\rm r0}(1+z)^4+\Omega^{\rm w}_{\rm m0}(1+z)^3+\Omega^{\rm w}_{\rm de0}(1+z)^{3(1+w_{\rm de})}\right]^{1/2},
\end{equation}
\begin{equation}\label{Hdot}
(1+z)\frac{dE^{\rm w}(z)}{dz}=\frac{3}{2}E^{\rm w}(z)\left\{1+\frac{ w_{\rm de}\Omega^{\rm w}_{\rm de0}(1+z)^{3(1+w_{\rm de})}+\frac{1}{3}\Omega^{\rm w}_{\rm r0}(1+z)^4}{\left[E^{\rm w}(z)\right]^2}\right\},
\end{equation}
where $E(z)=\text{H}(z)/\text{H}_{\rm 0}$ is the dimensionless expansion parameter. We use the superscript $w$ to identify that the density parameters are calculated within the $w$CDM cosmology.

The EoS parameter in $w$CDM model is commonly treated as a constant, but even for time dependent EoS parameters the evolution can become phantomic. Indeed, viscous fluid models produce a dynamical EoS. 

\subsection{Viscosity in the matter or radiation fluid}

Let us now assume a cosmological background composed by standard radiation fluid, dark energy given by cosmological constant $\Lambda$ and the matter component has bulk viscous pressure
\begin{equation}
T_{\left({\rm eff}\right)}^{\mu \nu}=T_{\left({\rm r}\right)}^{\mu \nu}+T_{\left({\rm vm}\right)}^{\mu \nu}+T_{\left({\rm \Lambda}\right)}^{\mu \nu},
\end{equation}
where the subscript ``{\rm vm}'' stands for viscous matter. The pressure of each component is given by
\begin{equation}
p_{\rm r}=\frac{\rho_{\rm r}}{3}, \hspace{1cm}p_{\rm vm}=-3\text{H} \xi_{\rm vm}, \hspace{1cm}p_{\rm de}=-\rho_{\rm de},
\end{equation}
where for the bulk viscous coefficient of the viscous matter we assume
\begin{equation}
\xi_{\rm vm}\equiv \xi_{\rm vm}(\rho_{\rm vm})=\xi_{\rm vm0}\left(\frac{\rho_{\rm vm}}{\rho_{\rm vm0}}\right)^{\nu}.
\end{equation}

From the above choice we note that for a constant bulk viscous coefficient ($\nu=0$) the pressure $p_{\rm vm}$ does not depend explicitly on the energy density $\rho_{\rm vm}$. Indeed we have $p_{\rm vm}\equiv p_{\rm vm}(\text{H})$ and the analysis of section 2 applies, i.e, for a constant bulk viscous coefficient there are viscous cosmologies which can not be distinguished from dark energy phantom models. Our task now is to understand the equivalence between viscous and phantom dark energy cosmologies if $\nu \neq 0$.

For a general value of the parameter $\nu$ the dynamics is given by the expression
\begin{equation}\label{Evm}
E^{\rm vm}(z)=\left[\Omega^{\rm vm}_{\rm r0}(1+z)^4+\Omega^{\rm vm}_{\rm vm}(z)+\Omega^{\rm vm}_{\rm de0}\right]^{1/2},
\end{equation}
where the density of the viscous matter has to be determined from the energy balance
\begin{equation}\label{Omegavmdot}
(1+z)\frac{d \Omega_{\rm vm}}{dz}-3\Omega_{\rm vm}+\tilde{\xi}_{\rm vm}E^{\rm vm}(z)\left(\frac{\Omega_{\rm vm}}{\Omega_{\rm vm0}}\right)^{\nu}=0 ,
\end{equation}
with
\begin{equation}
\tilde{\xi}_{\rm vm0}=\frac{24\pi G \xi_{\rm vm0}}{\text{H}_{0}}.
\end{equation}

Of course, the degeneracy problem means two cosmological models which result in the same expansion of the universe. Assume that we have two solutions which are named $E_1(z;\theta)$ and $E_2(z;\gamma)$, where
$E=\text{H}/\text{H}_0$, $\theta$ and ${\gamma}$ represent two different parameter sets. If they
are degenerate solutions,then $E_1(z;\theta)=E_2(z;\gamma)$. According to this
\begin{equation}\label{deg}
        E^{\rm w}(z)=E^{\rm vm}(z)
\end{equation}
should be satisfied. With the above equality we can map parameter-space of the matter viscous model into the parameter-space of $w$CDM model. Hence, we have
\begin{equation}
        \Omega_{\rm vm}(z)=\Omega_{\rm m0}^w(1+z)^3+(\Omega_{\rm r0}^w-\Omega_{\rm r0})(1+z)^4+\Omega_{\rm de0}^{\rm w}(1+z)^{-3\delta}-\Omega_{\rm de0},
\end{equation}
where $\delta =-(1+w_{\rm de})$. It is apparent that only if $\delta$ is positive, we can derive phantom $w$CDM model by the concept of model degeneracy. Substituting
the above equation of $\Omega_{\rm vm}(z)$ into Eq. (\ref{Omegavmdot}) we find
\begin{equation}\label{delta}
        (\Omega^{vm}_{\rm r0}-\Omega_{\rm r0}^{\rm w})(1+z)^4+3[\Omega_{\rm de0}^w(1+\delta)(1+z)^{-3\delta}-\Omega^{\rm vm}_{\rm de0}]=\tilde{\xi}_{\rm vm0}E^{\rm vm}(z)\left(\frac{\Omega_{\rm vm}(z)}{\Omega_{\rm vm0}}\right)^\nu.
\end{equation}

In what follows we will make two assumptions about the above equality. Firstly, since the density of radiation fluid can be directly obtained from the temperature of the CMB photons it is clear that this quantity is not model dependent and we can safely assume $\Omega^{\rm vm}_{\rm r0}=\Omega^{\rm w}_{\rm r0}$. 

Let us also assume that observations are able to establish the correct quantity of matter $\Omega_m$ independently of the model. This would imply $\Omega^{\rm w}_{\rm m0}=\Omega_{\rm vm0}$. As a consequence, the dark energy density parameter is also well determined for any cosmology therefore $\Omega^{\rm w}_{\rm de0}=\Omega^{\rm vm}_{\rm de0}$. 

With these assumptions we can simplify (\ref{delta}) to

\begin{equation}
3\Omega^w_{\rm de0}\left[(1+\delta)(1+z)^{-3\delta}-1\right]=\tilde{\xi_{\rm vm0}}E^{\rm vm}(z)\left(\frac{\Omega_{\rm vm}(z)}{\Omega_{\rm vm0}}\right)^\nu.
\label{delta1}
\end{equation}

If the above condition is satisfied we can write the matter viscous cosmology Eqs. (\ref{Evm}) and (\ref{Omegavmdot}) in terms of the parameters of the phantom dark energy case, i.e., we have to solve

\begin{equation}\label{dvm}
(1+z)\frac{d \Omega_{\rm vm}}{dz}-3\Omega_{\rm vm}+3\Omega_{\rm de0}\left[(1+\delta)(1+z)^{-3\delta}-1\right]=0 ,
\end{equation}
for the density $\Omega_{\rm vm}$.

Note that as a consequence of the second law of thermodynamics $\xi \geq 0$. Then, from Eq. (\ref{dvm}), positive values of $\tilde{\xi}$ should be mapped into positive $\delta$. 

In Eq. (\ref{dvm}) we re-express the viscous DM model in terms of the $w$CDM's parameters. It is worth noting that $\tilde{\xi}$ and $\nu$ are characterized now only by $\delta$ which means the dimension of parameter-space has been reduced by one degree. Actually, this means that the degeneracy does not depend on the specific choice of the viscosity.

We also consider the $\Lambda$CDM model but taking into account the viscous pressure in the radiation fluid. This is the most likely scenario that can occur in the universe since it is well known that neutrinos do have bulk viscous properties \cite{degroot}.

The pressure of each component is
\begin{equation}
p_{\rm vr}=\frac{\rho_{\rm vr}}{3}-3\text{H} \xi_{\rm vr}, \hspace{1cm}p_{\rm m}=0, \hspace{1cm}p_{\rm de}=-\rho_{\rm de},
\end{equation}
where for the bulk viscous coefficient of the viscous radiation we assume
\begin{equation}
\xi_{\rm vr}\equiv \xi_{\rm vr}(\rho_{\rm vr})=\xi_{\rm r0}\left(\frac{\rho_{\rm vr}}{\rho_{\rm vr0}}\right)^{\nu}.
\end{equation}

We assume again that the density of radiation fluid is not model dependent. Thus, $\Omega^{\rm vm}_{\rm r0}=\Omega^{\rm w}_{\rm r0}$.
Hence,
\begin{equation}
\Omega^{\rm w}_{\rm de0}\left[(4+3\delta)(1+z)^{-3\delta}-4\right]=\tilde{\xi}_{\rm vr0}E^{\rm vr}(z)\left(\frac{\Omega_{\rm vr}(z)}{\Omega_{\rm vr0}}\right)^\nu.
\label{delta2}
\end{equation}

The dynamics is given by the expression
\begin{equation}\label{Evr}
E^{\rm vr}(z)=\left[\Omega_{\rm r}(z)+\Omega_{\rm m}(1+z)^3+\Omega_{\rm de0}\right]^{1/2},
\end{equation}

where the density of the viscous radiation has to be determined from the energy balance
\begin{equation}\label{Omegavrdot}
(1+z)\frac{d \Omega_{\rm vr}}{dz}-4\Omega_{\rm vr}+\tilde{\xi}_{\rm vr0}E^{\rm vr}(z)\left(\frac{\Omega_{\rm vr}}{\Omega_{\rm vr0}}\right)^{\nu}=0 ,
\end{equation}
with
\begin{equation}
\tilde{\xi}_{\rm vr0}=\frac{24\pi G \xi_{\rm vr0}}{\text{H}_0} .
\end{equation}

If the above condition is satisfied we can write the radiation viscous cosmology Eqs. (\ref{Evr}) and (\ref{Omegavrdot}) in terms of the parameters of the phantom dark energy case. Then, we solve

\begin{equation}\label{dvr}
(1+z)\frac{d \Omega_{\rm vr}}{dz}-4\Omega_{\rm vr}+\Omega_{\rm de0}\left[(4+3\delta)(1+z)^{-3\delta}-4\right]=0.
\end{equation}




\subsection{Comparison with observations}

The background expansion determined by Eqs. (\ref{Evm}, \ref{Evr}) will be used in this section to place constraints on the model parameters. Our goal here is to illustrate that different data sets indeed favour a phantom scenario.

According to our criterion, the parameters characterizing viscous fluid models, i.e., $\nu$, $\xi$, can be replaced by the parameters of $w$CDM model as shown in Eqs. (\ref{dvm}, \ref{dvr}). We constrain the viscous fluid models in terms of the $w$CDM parameters, which means that we transfer the constraining effect of data-sets between cosmological models. 

We consider the recent SNIa data set UNION2.1 \cite{UNION21} which includes 580 data points. Using (\ref{Evm}) with $\Omega_{\rm m}$ determined by (\ref{dvr}) for our viscous matter model and (\ref{Evr}) with $\Omega_{\rm r}$ determined by (\ref{dvr}) for our viscous radiation model, we calculate the distance modulus as a function of the free parameters
\begin{equation}
\mu^{\rm th}(z)=5log\left[(1+z)c \int^{z}_{0}\frac{dz^{\prime}}{\text{H}(z^{\prime})}\right]+25.
\end{equation}

We also use the baryon acoustic oscillation (BAO) scale which is calculated by the $D_{\rm V}$ parameter
\begin{equation}
D_{\rm V}(z)=\left[(1+z)^2 D_{\rm A}^2(z)\frac{c z}{\text{H}(z)}\right]^{1/3},
\end{equation}
where $D_{\rm A}(z)$ is the angular-diameter distance. The values for $D_{\rm V}$ have been reported in the literature by several galaxy surveys. 
In our analysis we use the SDSS \cite{sdss}, WiggleZ \cite{wiggle} and the 6dFGRS \cite{6dfgrs} data.

The third data is CMB distance priors \cite{komatsu} which contains CMB-shift parameter $\mathcal{R}$, red-shift $z_{\ast}$ at last scattering, and the position of the observed CMB peak $l_1=220.8 \pm 0.7$ obtained by the WMAP project \cite{CMB} that is related to the angular scale $l_{\rm A}$ \cite{wu},
\begin{equation}
l_1=l_A\left[1-0.267\left(\frac{r}{0.3}\right)^{0.1}\right],
\end{equation}
where $r=\rho_{\rm r}(z_{ls})/\rho_{\rm m}(z_{\rm ls})$ and $z_{\rm ls}$ is the redshift at the last scattering. The acoustic scale is calculated as
\begin{equation}
l_{\rm A}=\pi\int^{z_{\rm ls}}_{0}\frac{dz}{\text{H}(z)}/\int^{\infty}_{z_{\rm ls}}\left(3+\frac{9}{4}\frac{\Omega_{\rm b}}{ z \Omega_{\rm r}}\right)^{-1/2}\frac{dz}{\text{H}(z)}.
\end{equation}

Observed Hubble parameter data-sets (OHD or H(z) data-sets which contains 29 measurement points) are also included \cite{Hz}. The determination of the value of Hubble parameter at different redshifts relies on the differential age method \cite{galage}, utilizing mainly early-type galaxies observed by projects like SDSS, and combining measurements of BAO peak using data released from projects like WiggleZ. 

\begin{figure}[!h]
\begin{center}
\subfigure{\includegraphics[width=0.4\textwidth]{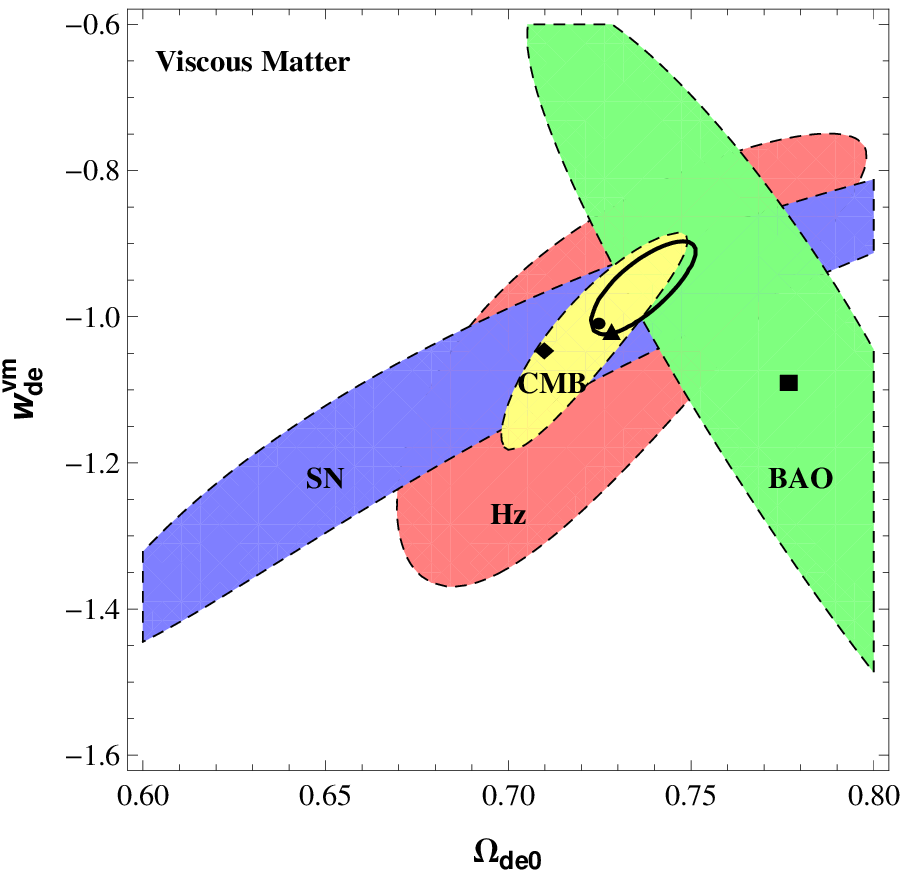}}
\subfigure{\includegraphics[width=0.4\textwidth]{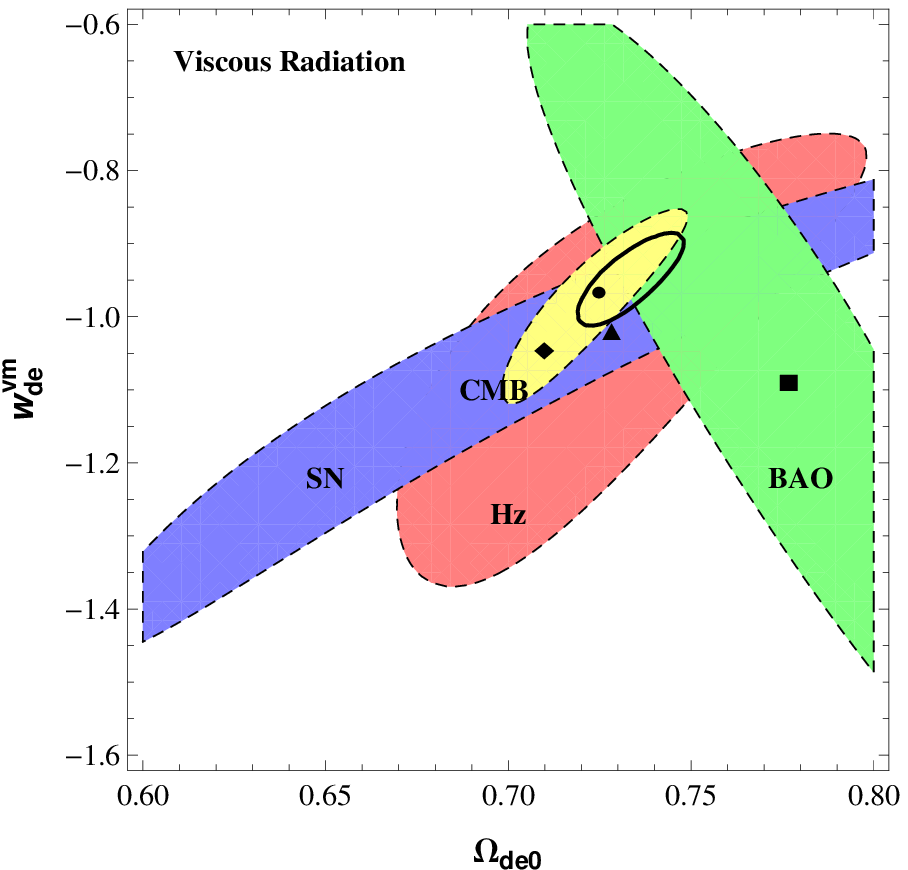}}
\label{fig1}	
\caption{Constraints on the matter viscous (left pannel) and radiation viscous (right pannel) models written in terms of the parameters of the phantom cosmology. We show $2 \sigma$ contours of confidence level for each data set separately. The SN data is shown in blue with best fit at the filled diamond. BAO in green with best at the filled square. H(z) data in red with best fir at filled triangle and CMB data in yellow with best fit at the filled circle.}
\end{center}
\end{figure}

In Fig.~\ref{fig1} we show constraints on the free parameters of the viscous matter (left panel) and viscous radiation (right panel) cosmology. The Hubble parameter has been fixed to $\text{H}_0=70~km\cdot s^{-1}\cdot Mpc^{-1}$, and radiation density at present is set as $\Omega_{\rm r0}h^2=4.31419\cdot 10^{-5}$ according to Planck's report ~\cite{Planck2013CP}. The dashed lines are the $2\sigma$ confidence level contours obtained from the likelihood function $\mathcal{L}\propto {\rm exp}(-\chi^{2}/2)$. The standard $\chi^{2}$ statistics,
\begin{equation}
\chi^2 (w_{\rm de0},\Omega_{\rm de0})=\sum^{N}_{i=1}\frac{(Q^{\rm th}_{i}-
Q^{\rm obs}_{i})^2}{\sigma^2_i},
\end{equation}
measures the goodness of the fit. For each data set, with $N$ data points, the theoretical value obtained for our model $Q^{\rm th}$ is confronted with the observation $Q^{\rm obs}$, where $\sigma^2_i$ is the error in each data point at $z_{i}$.


It is worth noting that the statistical results show no sensitivity to the specific choice of the coefficient of bulk viscosity. Indeed, our criterion (\ref{deg}) allowed us to write the dynamics in terms of only one parameter $\delta$ rather than $\tilde{\xi}_{\rm vm0}$ and $\nu$. Constraints on the magnitude of the viscosity would possible only if we fix the parameter $\nu$. Constraints on the dark matter viscosity can be found in Ref. \cite{viscousDM}.

\section{Conclusions}

Apart from the standard scenario where the cosmological constant $\Lambda$ drives the accelerated expansion, parametrizations for the pressure of the dark energy of the type $p_{\rm de}=w_{\rm de}\rho_{\rm de}$ are usually invoked. If one assumes $w_{\rm de}$ as a free parameter, the null energy condition imposes the prior $w_{\rm de} \geq -1$. For the particular value $w_{\rm de}=-1$, $\Lambda$ is recovered.  However, phantom models arise in cosmology when one relaxes the theoretical prior $w_{\rm de} \geq -1$. Interesting, as argued by the analysis presented in \cite{Phantomdata}, phantom dark energy is actually preferred by the current observational data sets. Even taking the more conservative side, one can at least say that available data is not able to rule out a dark energy equation of state in the regime $w_{\rm de} < -1$.

The main questions addressed in this work are the following. Is dark energy a fluid with an equation of state $w_{\rm de} < -1$? Is there some physical mechanism leading us to misinterpret the cosmological constant as a phantom fluid? We have argued that the existence of a cosmological non-equilibrium (bulk viscous) pressure in either the matter or the radiation components can explain such preference for values $w_{\rm de} < -1$. We investigated such possible degeneracy between a phantom cosmological model and viscous $\Lambda$CDM scenarios. 

We show that phantom DE cosmology can be realized by accepting the existence of bulk viscosity in the standard $\Lambda$CDM model. In section \ref{sectwo} we started our demonstration for coefficients of bulk viscosity of the form $\xi \equiv \xi(\text{H})$. Standard cosmology assumes that different fluids in the universe do no interact, i.e., each fluid obeys independently an equation for its energy conservation. Therefore the decomposition of the total energy conservation into the individual energy conservations is arbitrary. This means that a modified $\Lambda$CDM model where CDM has a small bulk viscous pressure  can not be differentiated from a $w$CDM model with $w=-1-\Pi/\rho_{\rm de}$ when only the background evolution is concerned. This same comparison holds for a $\Lambda$CDM model where radiation has a small bulk viscous pressure.

The cases where the coefficient of bulk viscosity $\xi_{\rm i}$ of some fluid {i} depends of its own density $\xi_{\rm i}\equiv \xi_{\rm i}(\rho_{\rm i})$ were discussed in section 3. In this case, the decomposition studied in section \ref{sectwo} is not arbitrary anymore. We study both matter viscous (with $p_{\rm m}\propto\xi_{\rm m}\equiv \xi_{\rm m}(\text{H})$) and radiation viscous (with $p_{\rm r}\propto\xi_{\rm r}\equiv \xi_{\rm r}(\text{H})$) cosmologies. Our strategy was to write such viscous cosmologies in terms of the $w$CDM parameters. This correspondence is achieved by requiring that the expansion in both models is indeed the same. 

We tested the combined parameter pair $\{w^{\rm vm}_{\rm de}/w^{\rm vr}_{\rm de},\Omega_{\rm de0}\}$ with multiple astrophysical data-sets which rely on background cosmological evolution. The constraining results support our suggestion, by illustrating that the phantom interpretation is allowed in viscous fluid models.
We have used observational data from Supernovae, BAO, CMB and H(z) to show that values $w_{\rm de}<-1$, which in our notation corresponds to $\delta > 0$, are preferred. This shows that phantom cosmology can be realized by adding an extra imperfect pressure from bulk viscosity to any component of the standard $\Lambda$CDM model without introducing scalar fields or handling modified gravity theories. 

We also have mentioned in the introduction that neutrinos do have bulk viscous properties. Depending on their masses (if they are massive), neutrinos undergo a transition from the relativistic to the non-relativistic regimes at some point during the cosmic history. Both viscous cosmologies studied here could be a first approximation to understand the cosmological dynamics of a viscous neutrino component. We leave for a future work to study the impact of bulk viscosity on the current bounds on the sum of the neutrino masses. 

The degeneracy pointed out in this study concerns only the background evolution. Also, we assumed that the different models have the same matter density. Indeed, observations of large scale structures seem place strong constraints on this parameter \cite{mreqWiggle}. Although this assumption restricts the space parameter of the viscous cosmology, it is a necessary assumption if we want to transfer the effects of $\xi$ into $w^{\rm w}_{\rm de}$ only. 

We have limited here to the background analysis. First order perturbations have not been included in this work. Since the viscous matter density will evolve in time differently with respect to the standard CDM case, it is expected that this assumption leads to  distict moment of the matter-radiation equality ($z_{eq}$) and the time of last scattering ($z_{lss}$). However, the matter density could be adjusted in viscous models in order to recover the same values for $z_{eq}$ and $z_{lss}$ as in the corresponding $w$CDM model. Thus, the full calculation of the spectrum of anisotropies in the cosmic microwave background and the matter power spectrum will be much more efficient in order break the degeneracy demonstrated here. We also hope to address this issue in a future work.

\section*{Acknowledgement}

HV thanks CNPq for financial support. This work is partly supported by Natural Science Foundation of China under Grant Nos.11075078 and 10675062, and by the project of knowledge Innovation Program (PKIP) of Chinese Academy of Sciences (CAS) under the grant No. KJCX2.YW.W10 through the KITPC astrophysics and cosmology program where we have initiated the present research.

\end{document}